Exploring Percolative Landscapes: Infinite Cascades of Geometric Phase Transitions

P. N. Timonin[1], Gennady Y. Chitov[2]

(1) Physics Research Institute, Southern Federal University,

344090, Stachki 194, Rostov-on-Don, Russia, pntim@live.ru

(2) Department of Physics, Laurentian University
Sudbury, Ontario P3E 2C6, Canada, gchitov@laurentian.ca

The evolution of many kinetic processes in 1+1 (space-time) dimensions results in 2d directed percolative landscapes. The active phases of these models possess numerous hidden geometric orders characterized by various types of large-scale and/or coarse-grained percolative backbones that we define. For the patterns originated in the classical directed percolation (DP) and contact process (CP) we show from the Monte-Carlo simulation data that these percolative backbones emerge at specific critical points as a result of continuous phase transitions. These geometric transitions belong to the DP universality class and their nonlocal order parameters are the capacities of corresponding backbones. The multitude of conceivable percolative backbones implies the existence of infinite cascades of such geometric transitions in the kinetic processes considered. We present simple arguments to support the conjecture that such cascades of transitions is a generic feature of percolation as well as of many other transitions with nonlocal order parameters.

1. Introduction

The cornerstones of modern civilization are various types of networks: telecommunications, electric power supply, Internet, warehouse logistics, to name a few. Their proper functioning is of great importance and many studies have been devoted to their robustness against various kinds of damages [1, 2]. Yet there are other aspects of the network quality – their abilities for restoration, repairing, renovation and upgrades. To asses these abilities, we should know the details of the network structure which are responsible for these features. Also these details should be implemented in the constructions of new networks. A useful property of a network can be the existence in it of some connected subset spreading through the whole network. We will call this subset the *percolative backbone.*

For example, a conceivable way to modernize the transportation or communication networks could be by placing more powerful transmitters on the backbone nodes and discarding the obsolete ones on the rest of the nodes. Also, in the process of damaged network restoration one can link the remnant backbone nodes to quickly restore the supply on a large territory. In case when the network is a planar percolation cluster on a lattice, the percolative backbone can be comprised of the cluster nodes belonging to a certain sublattice. Another conceivable way of the network upgrade is to merge groups of several nearby nodes into a single hub. These hubs can comprise the backbone made out of the ``coarse-grained" original network. Percolative properties of clustered and coarse-grained networks have been intensively studied in the recent literature [2–5].

In this paper we define the percolative backbone as a subset of the original percolation cluster made out of its nodes or hubs, such that the subset itself forms a percolative cluster. The percolation in the backbone is defined with respect to the bonds between, e.g., nearest or next-nearest, etc., neighbors of the backbone cluster. In general, these new bonds do not correspond to their counterparts in the original network. This quite formal definition becomes more transparent after we apply it to several models analyzed in the subsequent sections of the paper.

The majority of the real networks emerge as a result of some stochastic process spreading in space according to its landscape, population, resources distribution, etc. Whether the networks are renewable or reparable in the sense discussed above is not known a priori and can be addressed by studying putative percolative backbones defined in those networks. For the model networks these properties can be established via analysis of the Monte Carlo (MC) simulations of the kinetic processes. In this paper we use the MC technique to explore the 2*d* (space-time) percolation patterns (networks) emerging in the directed percolation (DP), contact process (CP) [6, 7] and the replication process introduced in Ref. [8]. We show that under variations of control parameters the percolation patterns of these processes undergo a series of geometrical phase transitions signaling emergence of various percolative backbones. We present arguments that such cascades are a generic feature of percolation as well as of other transitions with nonlocal order parameters.

The paper is organized as follows: In Section 2 the MC simulations of the directed percolation are presented. The results reveal a set of the backbone transitions in the directed percolation clusters. There we give the formal analytic definition of the backbone order parameter and the recurrence scheme of its calculation from the MC data. In Sections 3, 4 the analogous results for the lattice version of the contact process and for the replication process defined in Ref. [8] are presented. In Section 5 we discuss the generalization of the present results for other types of percolation models. The final Section 6 is devoted to conclusions and discussion of the relation between the geometrical transitions we found and the transitions in other models with nonlocal order parameters.

## 2. Directed percolation

The directed percolation on 2*d* (space-time) lattice shown in Fig. 1 can be considered as kinetic process. Setting the steps in the temporal or spatial directions equal to unity, the spacing of the DP lattice in Fig. 1 is a = √2. Each site can be in one of two states - wet or dry (filled-empty, infected- healthy etc., in other contexts). In the DP variant called "bond DP" (BDP) at each time step the percolative bonds are placed randomly with probability *p* between nearest neighbor sites in the columns *t* and *t*+1. If such bond connects a wet site at column *t* with its neighbor at column *t*+1 then the last also becomes wet, otherwise it stays dry. There can be two outcomes of this time evolution starting from some initial configuration of wet sites at *t* = 0 – either the wet sites become extinct or they persist for infinite times. The first scenario takes place for $p < p_{BDP} \approx 0.6447$ and it corresponds to the absorbing phase, while the infinite proliferation of wet sites appears for $p > p_{BDP}$ resulting in the active or percolating phase [6, 7] .

The main tool for the studies of the DP-type kinetic processes is the Monte-Carlo simulations. They numerically mimic stochastic evolution based on the model's transfer probabilities defined as the probability of possible configurations at time *t*+1 given the configuration at time *t*. We denote it as $P(n_{i,t+1}|n_{i-1,t},n_{i+1,t})$. Here all sites are endowed with occupation numbers (a.k.a. lattice gas parameters) $n_{i,t}$ = 0,1; where 1 corresponds to wet (filled) sites and 0 is ascribed to dry (empty) ones. For the BDP the transfer probabilities are

$$P(1|0,1) = P(1|1,0) = p , P(1|1,1) = p(2-p), P(1|0,0) = 0; P(0|a,b) = 1 - P(1|a,b).$$

Note that the probability of site to become wet when it has two wet ancestors is the probability to have at least one bond attached to the site, i. e. 1 - (1-*p*)$^2$ = *p*(2-*p*). In this work we implement the Monte-Carlo simulations with parallel update, in which the configurations of all sites are updated simultaneously in one time step. This very simple numerical procedure results in raw data whence various (BDP) percolating patterns (networks of wet sites with $n_{i,t}$ = 1) can be revealed. Note that the ordinary absorbing- active transition can be formally described with the local order parameter $\rho(t) = \left\langle \sum_i n_{i,t} \right\rangle / N = \left\langle n_{i,t} \right\rangle$ such that $\rho(\infty) = 0$ for $p < p_{BDP}$ and $\rho(\infty) > 0$ for $p > p_{BDP}$. In spite of the local form of $\rho(t)$ it also describes the nonlocal ordering in active phase giving simultaneously the probability of the existence of connected path of active sites in the [0, t] time interval. This is the consequence of the fact that every filled site has at least one ancestor in preceding time step as follows from transfer probabilities of the model.

For every trial with $p > p_{BDP}$ we get chaotically looking percolative patterns which appear to have only two features in common: wet sites always spread to infinite time and the number of wet sites at large times is approximately the same at all trials for a given value of p. In other respects these patterns are seemingly disordered and devoid any structure for all $p > p_{BDP}$. Yet, motivated by the problems of repairing and modernizing of percolative networks, one can ask if appearing percolative patterns possess inner coarse-grained structures over which some sort of percolation is still possible and at what $p > p_{BDP}$ they exist.

However, it is well known for transitions with local order parameters that various possible coarse-graining implemented in the sense of the Kadanoff-Wilson renormalization group cannot yield new critical points. In other words, the coarse-grained versions of the local order parameter should appear simultaneously with the original one. Yet it is not quite evident that the same is true for nonlocal ordering. Implementing various coarse-grained backbone schemes as nonlocal order parameters, we can find if they appear simultaneously with the local order $\rho(\infty)$ at $p = p_{BDP}$ or this take place at some larger p. The surprising new result, we intend to prove in the following, is that some backbone schemes engendered cascades of new phase transitions at $p > p_{BDP}$. This means that the percolation patterns in the BDP model possess actually many types of the intrinsic geometrical structures that emerge subsequently with variation of p.

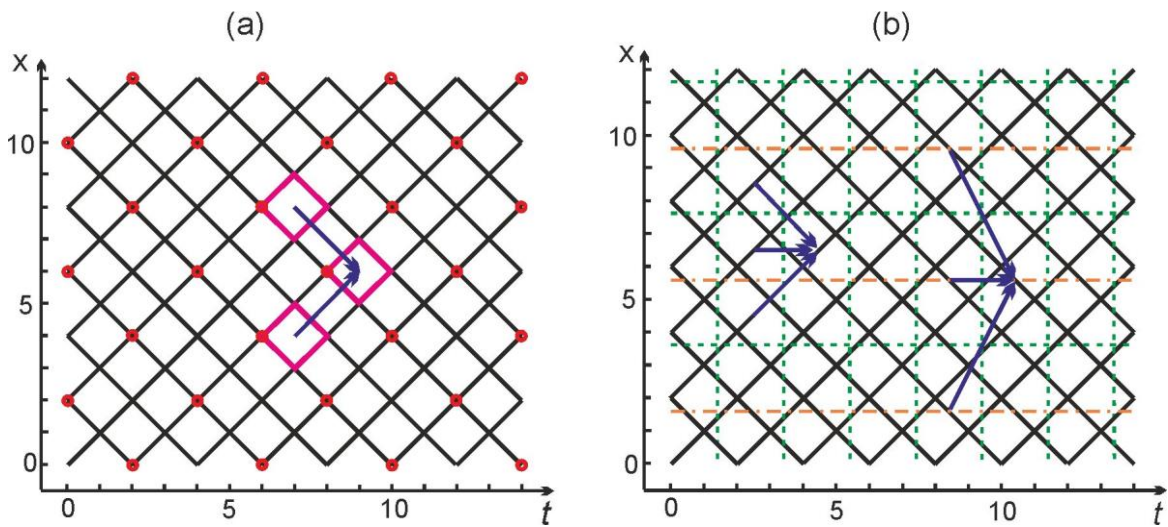

Fig.1. (Color online) *Examples of coarse-grained lattices (backbones) for the BDP process.* The original tilted square lattice for the BDP process with spacing $a = \sqrt{2}$ is shown by solid lines on both panels. (a) The tilted square sublattice with the spacing $2a$ and the sites denoted by red (grey) circles. Those circles also indicate the leftmost nodes of plaquettes (magenta [light-grey] squares) discussed in the text. (b) Division of the original lattice into 4-site cells (dashed lines) and into 2-site cells (dashed and dot-dashed lines). Colored arrows show the bonds for percolation on resulting coarse-grained lattices.

To reveal various percolative backbones let us first check whether the conventional percolative phase on the original tilted square BDP lattice with spacing $a$ (cf. Fig. 1) also contains a percolative cluster of the wet sites on the sublattice with spacing $2a$ shown in Fig. 1(a) with red circles. We consider sets of the sublattice sites belonging to the original pattern of directed percolation to determine whether there are paths connecting the nearest neighbors in this set, such that these paths traverse the whole sample in time direction thus forming the sublattice backbone of this particular percolation pattern. For such paths one can easily calculate the number of sites in the backbone at arbitrary t via a recursive numerical procedure. First we need to find connected backbone sites (CBS) at *t* = 2, i. e., those having nearest neighbors from the set at *t* = 0, then we search for CBS at *t* = 4 having nearest neighbors in CBS at *t* = 2, and so on. In doing so over many trials we can obtain at every even time step *t* the average number $N_{CBS}(t)$ of sites in the backbone connected by the nearest-neighbor paths to *t* = 0.

Analytical representation of $N_{CBS}(t)$ is readily given as an average of the order parameter operator

$$2N_{CBS}(t)/N = \langle \mathcal{P}_{i,t} \rangle \equiv P(t), \quad \mathcal{P}_{i,t} = \sum_{\sigma} O_{i,t}(\sigma) \qquad (1)$$

$$O_{i,t}(\sigma) = n_{2i,2t} \prod_{\tau=1}^{t-1} n_{2\left(i+\sum_{k=\tau}^{t-1}\sigma_k\right),2\tau} \qquad (2)$$

Here all auxiliary parameters **σ** can admit two values $\sigma_k = \pm 1$ and the angular brackets denote the averaging with the distribution function

$$W_{BDP} = \prod_{i,t} P(n_{i,t+1}|n_{i-1,t}, n_{i+1,t}) \equiv e^{-H_{BDP}}/Z, \quad H_{BDP} = \sum_{i,t} H_{i,t}$$

$$H_{i,t} = n_{i,t+1}\left[\ln\frac{p}{2-p}n_{i+1,t}n_{i-1,t} + \ln\frac{1-p}{p}(n_{i+1,t}+n_{i-1,t}) + i\psi(1-n_{i+1,t})(1-n_{i-1,t})\right] - 2\ln(1-p)n_{i,t} \quad (3)$$

Here the auxiliary variable of integration $\psi$ is introduced to enforce the model rule $(0,0) \to 0$. For more details on this formalism, see [8].

$W_{BDP}$ gives the probability of every configuration in BDP process for given initial values of $n_{i,0}$. Each set of **σ** defines the nearest-neighbors path on the sublattice connecting the site with coordinates (2*i*, 2*t*) to some sublattice site at *t* = 0. Operator $O_{i,t}(\sigma)$ is equal to 1 if the path belong to the percolation cluster of BDP process and zero otherwise. Thus its average in Eq. (1) is the probability that site (2*i*, 2*t*) belongs to the sublattice backbone. For cyclic boundary conditions in space direction the order parameter *P*(*t*) in Eq. (1) does not depend on the site index 2*i*.

It follows from Eq. (2)

$$O_{i,t+1}(\sigma) = n_{2i,2t+2}\left(O_{i-1,t}(\sigma')\delta_{\sigma_t,-1} + O_{i+1,t}(\sigma')\delta_{\sigma_t,1}\right) \qquad (4)$$

where **σ'** is **σ** without $\sigma_t$. Summing this equation over **σ** we get

$$\mathcal{P}_{i,t+1} = n_{2i,2t+2}\left(\mathcal{P}_{i-1,t} + \mathcal{P}_{i+1,t}\right). \qquad (5)$$

Thus for any given configuration of $n_{i,t}$, operator $\mathcal{P}_{i,t}$ can be calculated iteratively starting from $\mathcal{P}_{i,0} = n_{2i,0}$. Just this procedure is implemented in our MC calculations while the average $\langle \mathcal{P}_{i,t} \rangle = P(t)$ is obtained via the averaging over MC trials. We used chains with *N* = 20000 sites with *T* = 2000 time steps and averaged raw data over 200 trials. The cyclic boundary conditions are imposed and initial state is fully occupied. The results are presented in Fig. 2. To determine the approximate transition point we find first at which *p* the *P*(*t*) relaxation curve in Fig. 2(a) is better described by the power low $c/t^\alpha$ shown as dashed straight line in this figure. Thus we find the approximate $\alpha$ and $p_c$ such that curves with $p > p_c$ tend upward in Fig. 2(a) indicating the onset of long-range order $P(\infty) \neq 0$. Then we use variational procedure to collapse all *P*(*t*) curves in Fig. 2(a) into single scaling function $t^\alpha P(t) = R\left[t|p-p_c|^{\nu_\|}\right]$ to find the indices and more precise $p_c$ starting with previously found $\alpha$, $p_c$ and trying various initial values of $\nu_\|$. The resulting scaling function with $p_c$ = 0.6635, $\alpha$ = 0.16 and $\nu_\|$ = 1.73 is shown in Fig. 2(b). Thus we can conclude that the stable BDP percolation patterns undergo the second order phase transition at $p_c$ in which they acquire the infinite 'sublattice backbone'. The scaling indexes that we found for this transition are those of the DP universality class. This is just what can be expected from the Janssen-Grassberger conjecture on kinetic transitions in models with Ising-like variables and without conservation laws [6]. However, the original conjecture was made for absorbing-active phase transitions and our results show that it can be expanded to the active-active transitions in such models, see Section 3 and 4.

Similar analysis can be carried out for other type of backbones. In particular, we can easil y modify the previous procedure to deal with the backbone formed by a sublattice with spacing 4$a$ and sites located at the points (4$i$, 4$t$). We just need to substitute 2 → 4 in the subscripts entering equation (2) to obtain the string operator, and then, the order parameter (1) for such a sublattice. To treat the backbones obtained by coarse graining of the original lattice, (cf. 2-site and 4-site cells shown in Fig. 1(b)), we introduce the occupation number of the cells $\nu_{j,\tau} = 0,1$ as

$$\nu_{j,\tau} = \vartheta\left(\sum_{i,t\in c_{j,\tau}} n_{i,t} - f\right), \qquad (6)$$

where $\vartheta$ is the Heaviside step function defined such that $\vartheta(x \geq 0) = 1$ and the summation runs over all sites in the cell $c_{j,\tau}$.

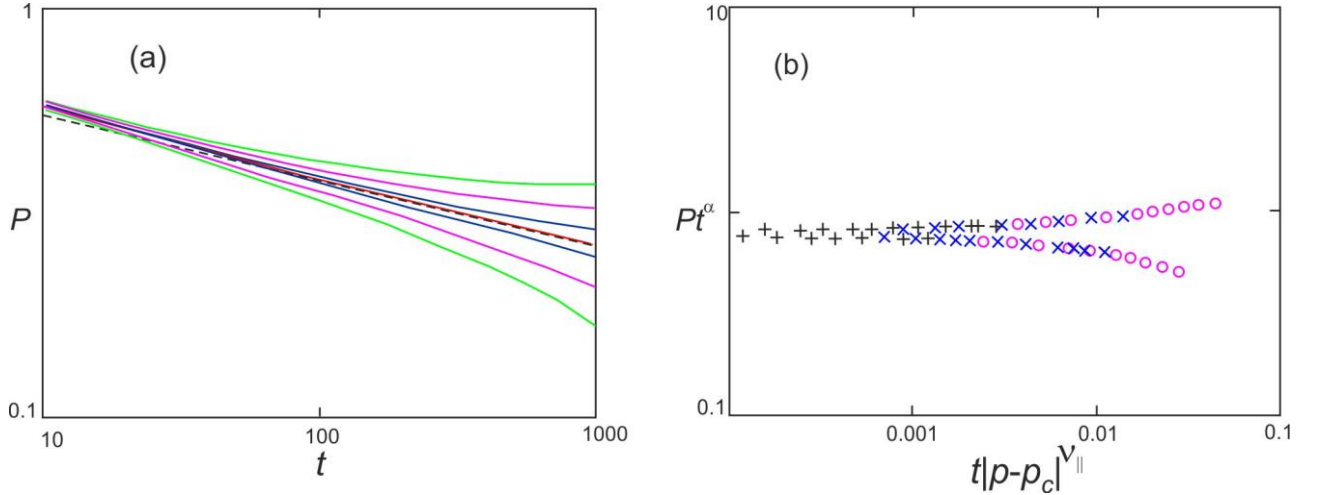

Fig. 2. (Color online) *BDP on the* 2*a sublattice.* (a) MC simulations of relaxation of the BDP sublattice percolation capacity *P(t)* for series of *p* near $p_c \approx 0.6635$, from top to bottom: *p=0.6665, 0.665, 0.664, 0.6635, 0.663, 0.662, 0.661.* The dashed line corresponds to the power law *0.8/ $t^\alpha$*, with $\alpha$ = 0.16. (b) Collapse of the curves from (a) onto a single scaling function. Fitting gives the values of $\nu_\| \approx 1.731$ and $p_c$ = 0.6635.

This new "renormalized" occupation number on the coarse-grained lattice depends on the overall cell filling *f*, and different choices of parameter *f* correspond to different types of backbones. We can as well choose different bonds for the cell percolation. For instance, as one can see from Fig. 1(b), we can allow percolation from three ancestors (nearest and next-nearest) or ban percolation from the nearest ancestor thus eliminating the horizontal bond. So we obtain different types of backbones. Note that in the case of percolation over a sublattice or via plaquettes, only two nearest ancestors are present on the renormalized lattice, cf. Fig. 1(a), so the above comments do not apply for the latter cases.

We ran simulations and scaling analyses explicitly for several backbones appearing in the percolative patterns for this model using *N*=20000, *T*=2000. The results are summarized in Table 1. The appearance of a new percolative backbone at the specific critical value $p_c$ constitutes a genuine second order phase transition with specific nonlocal order parameter (capacity of corresponding backbone) similar to the one defined by equations (2), (3). The critical indices we obtained from the collapse of the appropriate scaling curves for the critical points presented in Table 1 indicate that all these geometric phase transitions belong to the DP universality class. This can be expected for the transitions with scalar site variables and without conservation laws [6, 7].

Table 1. Critical points of geometric phase transitions where different backbones appear. The parameter *h* designate the presence (*h* = 1) or absence (*h* = 0) of horizontal bonds in the percolative backbone.

| Sublattices | | 2-site hubs 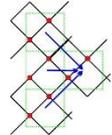 | | | Plaquettes 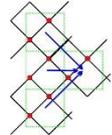 | | 4-site hubs 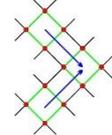 | | |
|---|---|---|---|---|---|---|---|---|---|
| Spacing | $p_c$ | *f* | *h* | $p_c$ | *f* | | *f* | *h* | $p_c$ |
| 2a | 0.663(5) | 1 | 1 | $p_{BDP}$=0.6447 | 1 | $p_{BDP}$=0.6447 | 2 | 1 | $p_{BDP}$=0.6447 |
| 4a | 0.677(6) | 1 | 0 | 0.646(7) | 2 | 0.646(5) | 2 | 0 | 0.646(4) |
| | | 2 | 1 | 0.656(2) | 3 | 0.671(2) | 3 | 1 | 0.657(4) |
| | | 2 | 0 | 0.696(3) | 4 | 0.742(5) | 3 | 0 | 0.682(7) |
| | | | | | | | 4 | 1 | 0.710(2) |
| | | | | | | | 4 | 0 | 0.759(5) |

Note that our schemes for the construction of backbones resemble the first steps of the Kadanoff-Wilson renormalization group approach devised for elimination of the short-range fluctuations of local order parameter (the density of active sites, $\rho = \langle n_{i,t} \rangle \big|_{t \to \infty}$ in our case). Actually only some of those order parameters are the coarse-grained versions of $\rho$ with transitions at the same critical point $p_{BDP}$, see the values of $p_c$ for the plaquette, 2-site and 4-site hubs backbones with *f* = 1 and *h* = 1 in the first line of Table 1. This implies that the order parameters of those backbones are not independent and the differences in their definitions amount to the coarse graining (rescaling) of $\rho$ in the sense of Kadanoff, thus do not result in shifts of the critical point.

The most remarkable feature of the model stemming from nonlocal nature of its ordering is that the other coarse-grained definitions of (nonlocal) order parameters give the percolation thresholds distinct from the critical value $p_{BDP}$, so we are dealing with a genuine cascade of new phase transitions. The distinct critical points are not finite-size artefacts: we have explicitly checked by doing simulations in larger samples and with longer times that only last digits of the values of $p_c$ given in Table 1 are affected. For example, the simulations of the chain with *N* = 60000 sites and *T* = 6000 time steps and with averaging over 200 trials yield for the 2a sublattice backbone the critical value $p_c$ = 0.6632(2) instead of $p_c$ = 0.663(5) for the case *N* = 20000 sites and *T* = 2000, cf. Table 1 and Fig. 2. So in larger samples transition points differ insignificantly from those presented in Table 1.

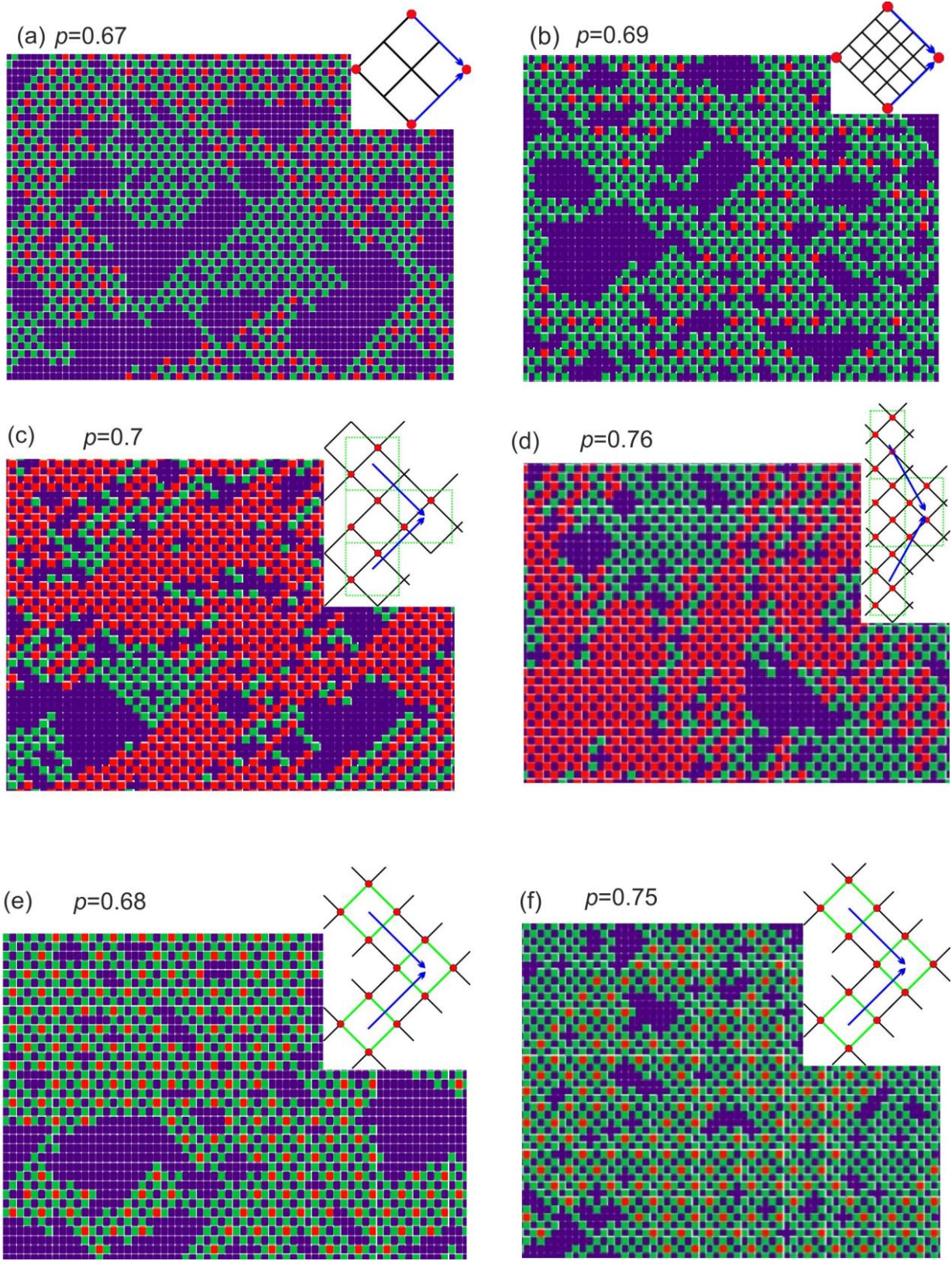

Fig. 3. (Color online) *BDP patterns*. The fragments of steady patterns of BDP process (green [light grey] squares) with backbones (red [grey] squares) and empty sites (blue [black] squares). (a) $p$ = 0.67, 2$a$ sublattice backbone, (b) $p$ = 0.68, 4$a$ sublattice backbone (c) $p$ = 0.68, 2-site cell backbone with next-nearest bonds, (d) $p$ = 0.76, 4-site cell backbone with next-nearest bonds, (e) $p$ = 0.68, plaquette backbone, $f$ = 3, red squares are centers of percolative plaquettes, (f) $p$ = 0.75, plaquette backbone, $f$ = 4.

We should also note that the backbones with the 2-site and 4-site hubs and $f = 1$, $h = 0$ appear simultaneously (within our precision) with the $f = 2$ plaquette percolative backbone at $p_c \approx 0.646$. For a more visual presentation of our results, several percolative patterns obtained by direct MC simulations along with corresponding schematic pictures of their backbones are shown in Fig. 3.

It should be obvious at this point, that one can construct in this way many other percolative backbones by varying sublattices, coarse-grained cells and/or percolative bonds. We conjecture that in the BDP model there is an infinite cascade of geometric phase transitions where various new percolative backbones emerge.

### 3. Contact process

Now we address another popular and well-studied kinetic model, namely, the contact process (CP). This model is used, e.g. to describe spreading of plant infections or plant population via dissemination of seeds [6, 7]. The model has two parameters $p$ and $q$, $p$ is the probability for an infected site to stay infected while the probability of a healthy one to become infected is proportional to $q$ and to the number of infected nearest neighbors. Similarly to the above analysis of the DP model we consider here the 2$d$ contact process on a chain with discrete time steps and parallel update using occupation numbers $n_{i,t}$. For a change we assume now that $n_{i,t} = 1$ corresponds to an infected site. The CP model's transfer probabilities are

$$P(1|0,0,1) = P(1|1,0,0) = q/2, \; P(1|1,0,1) = q, \; P(1|*,1,*) = p, \; P(1|0,0,0) = 0; \qquad (7)$$
$$P(0|a,b,c) = 1 - P(1|a,b,c).$$

The process undergoes absorbing-active phase transition of DP class [6, 7] at

$$p_{CP}(q) \approx 1 - aq - bq^3, \; a = 0.3, \; b = 0.17, \qquad (8)$$

with the order parameter $\rho(t) = \langle n_{i,t} \rangle$ such that $\rho(\infty) = 0$ at $p < p_{CP}(q)$ and $\rho(\infty) \neq 0$ at $p > p_{CP}(q)$. The critical line (8) shown in Fig. 4 is our fit of the MC simulation data. The CP patterns on the square time-space lattice represent 2$d$ percolative clusters in the temporal direction since all infected (survived) sites have at least one ancestor at the preceding time step similarly to the BDP process.

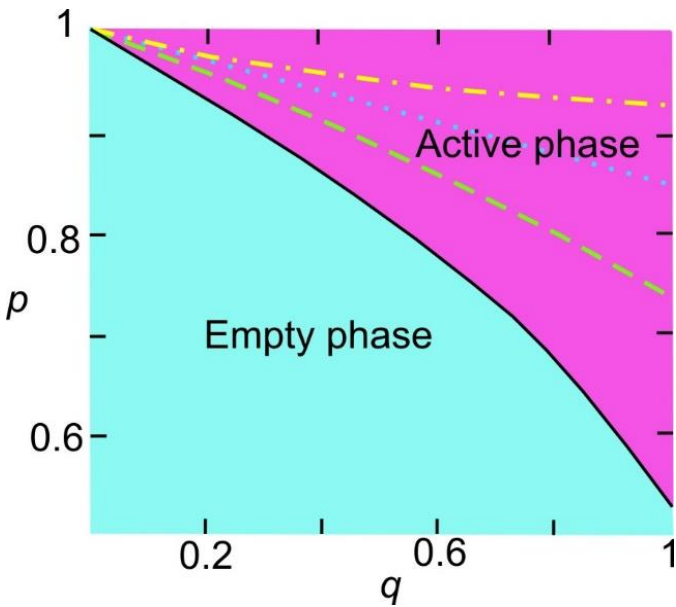

Fig. 4. (Color online) *Phase diagram of contact process*. The bold black curve (8) is the boundary of the absorbing-active phase transition. The boundaries for appearance of backbones with $h = 0$ are also shown. The sublattice backbone exists above dashed line, 2-site hub backbone – above dotted line and 4-site hub backbone with $f = 4$ – above dashed-dotted line.

We perform MC simulations to reveal various backbones in the percolative patterns. We consider sublattices with (*2i, 2t*) sites, 2-site cells of (*2i, t*), (*2i+1, t*) sites and 4-site cells of (*2i, 2t*), (*2i+1, 2t*), (*2i, 2t+1*), (*2i+1, 2t+1*) sites. We used the chains with $N$ = 20000 sites, $T$ = 2000 time steps and averaged over 200 trials. The cyclic boundary conditions are imposed and initial state is fully occupied.

Table 2. The transition points of contact process at $q = 1$ for several backbones.

| Sublattice | | 2-site hubs, $f = 2$ | | 4-site hubs | | |
|---|---|---|---|---|---|---|
| $h$ | $p_c(q=1)$ | $h$ | $p_c(q=1)$ | $f$ | $h$ | $p_c(q=1)$ |
| 1 | 0.603(5) | 1 | 0.732(7) | 2 | 1 | 0.528(3) = $p_{CP}(1)$ |
| 0 | 0.727(7) | 0 | 0.840(3) | 2 | 0 | 0.581(3) |
| | | | | 3 | 1 | 0.636(5) |
| | | | | 3 | 0 | 0.722(3) |
| | | | | 4 | 1 | 0.856(4) |
| | | | | 4 | 0 | 0.916(5) |

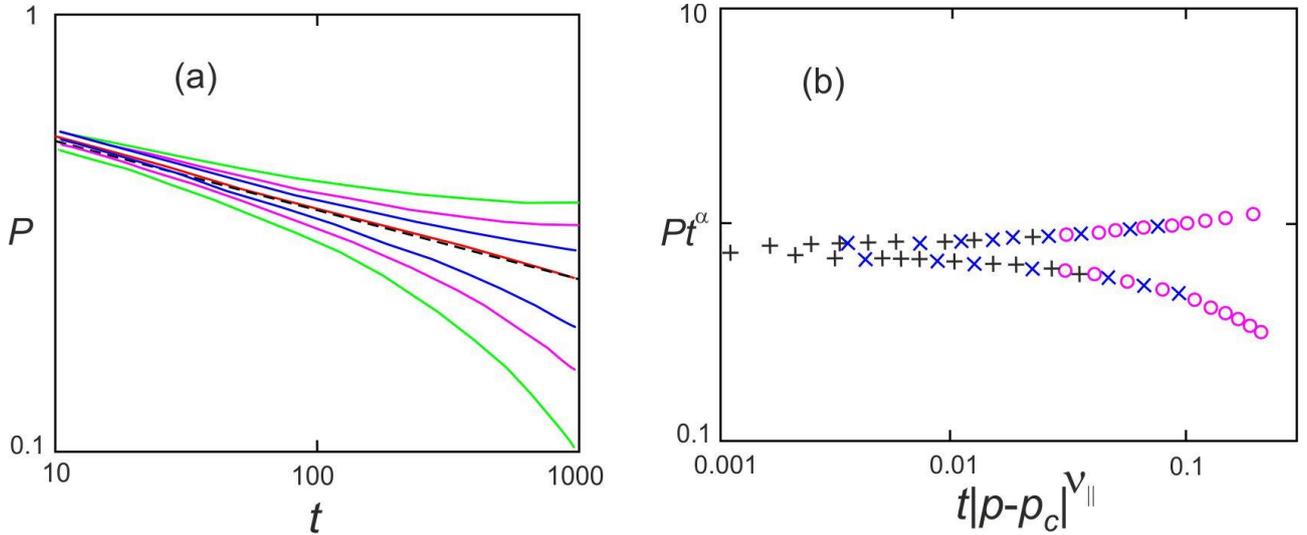

Fig. 5. (Color online) (a) MC simulations of relaxation of the CP sublattice percolation capacity $P(t)$ with $h = 0$ for $q = 1$ and series of $p$ near $p_c \approx 0.73$; from top to bottom: $p$ = 0.735, 0.732, 0.73, 0.7277, 0.725, 0.723, 0.72. Dashed line corresponds to power law $0.75/t^\alpha$, with $\alpha = 0.16$. (b) Collapse of the curves from (a) onto a single scaling function. Fitting gives the values of $\nu_\parallel = 1.735$ and $p_c = 0.7277$.

Similarly to the previously analyzed BDP model, the corresponding backbones also appear as a result of continuous geometric phase transitions. The critical lines for some of those transitions on the $p$-$q$ plane are shown in Fig. 4 The critical values of parameter $p$ at the ends of these lines ($q = 1$) are listed in Table 2.

The CP distribution function and analytical expressions for the order parameters of various backbones can be obtained similarly to those of the BDP, cf. equations (2)-(5). Direct MC simulations of the temporal relaxation of backbone capacities demonstrate the validity of scaling at all geometrical phase transitions. The values of the critical indices $\alpha = 0.16$ and $\nu_\parallel \approx 1.735$ imply that the transitions are of the DP universality class, see Fig. 5. The percolative patterns of CP with different backbones are shown in Fig. 6. Similarly to our previous conjecture about the BDP model, we infer that the various backbones we explicitly found and analyzed for the contact process are in fact only a small set of the infinite cascade of geometric phase transitions associated with the emergence of more and more new percolative backbones.

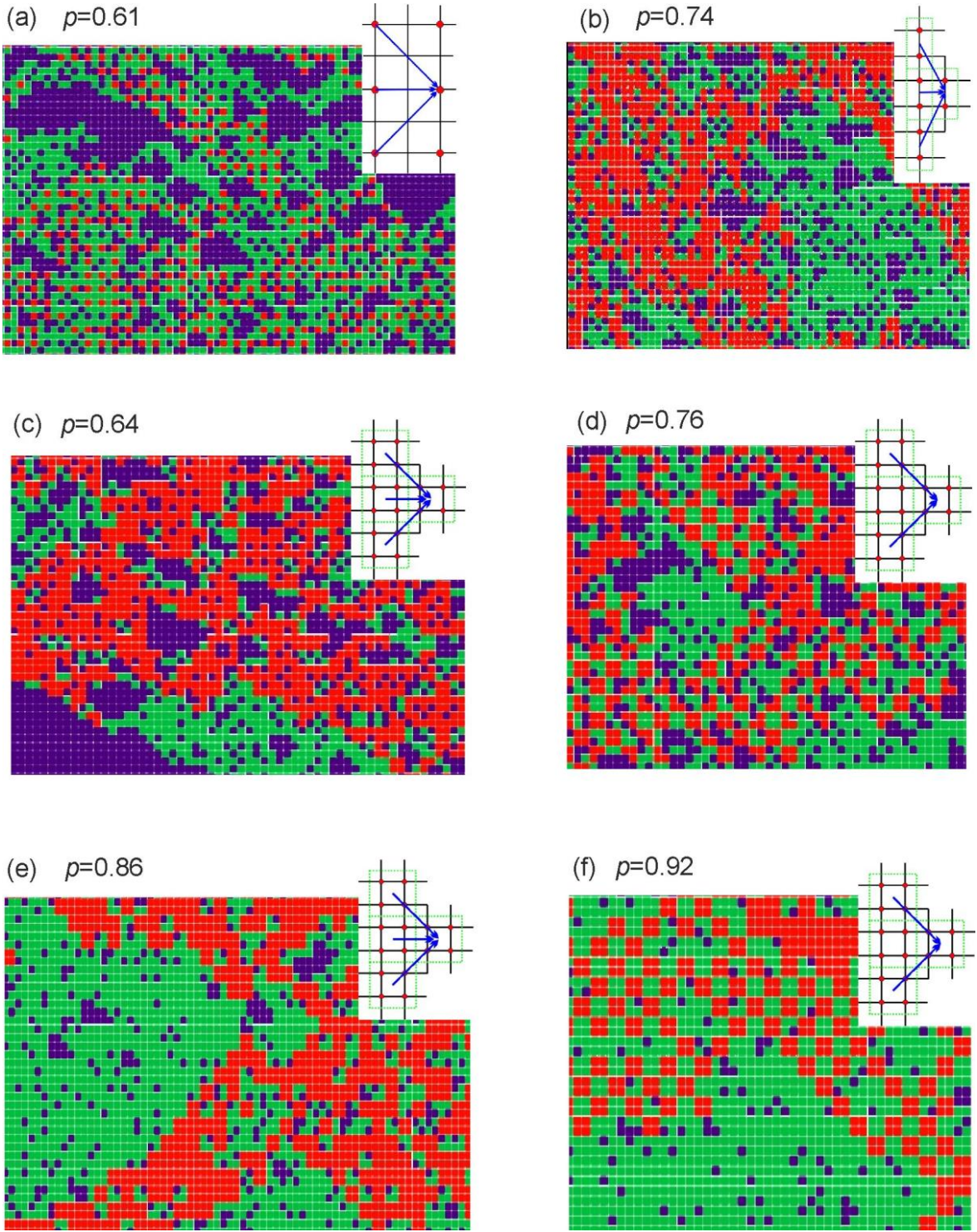

Fig. 6. (Color online) *Percolative patterns of contact process.* The fragments of steady patterns of contact process at $q = 1$ (green [light grey] squares) with backbones (red [grey] squares) and empty sites (blue [black] squares). (a) $p = 0.61$, $2a$ sublattice backbone, (b) $p = 0.74$, 2-site cell backbone with nearest and next-nearest bonds, $f = 2$, (c) $p = 0.64$, 4-site cell backbone with nearest and next-nearest bonds, $f = 3$, (d) $p = 0.76$, 4-site cell backbone with next-nearest bonds, $f = 3$, (e) $p = 0.86$, 4-site cell backbone with nearest and next-nearest bonds, $f = 4$, (f) $p = 0.92$, 4-site cell backbone with next-nearest bonds, $f = 4$.

## 4. Replication process

A variant of the kinetic replication process was introduced in [8]. Its transfer probabilities depend on the two parameters *p* and *q*. In the limiting cases $p \to 0$ and $p \to 1$ this process becomes BDP or CP, respectively. It was shown in [8] that besides standard absorbing-active transition there is another phase transition between two active phases. The unusual second active phase possesses a subtle percolative order. The backbone of that percolation pattern is made out of the active-dead nearest-neighbor spatial pairs unified in a connected network. Due to the analogy of that short-range spatial order to antiferromagnetic dipoles, we called it the antiferromagnetic percolative phase. In view of the relation of the model to the BDP and CP models, it is no surprise that other geometrical transitions considered above are also present in this case. In particular, our MC simulations revealed geometrical phase transitions inside the conventional active phase of the model associated with appearance of percolation via sublattices and/or coarse-grained lattices made out of various hubs. We will not give the technical details on these phases of the replication model, since they hardly give us more conceptual insights with respect to what has been presented so far. The key point is that the active phase of the model is not plain, and it also contains a cascade of geometric transitions due to the emergence of various percolative backbones.

## 5. Generalizations

We should note that many more percolative backbones could be considered in addition to those we have analyzed above. For instance, one can change the sum $\sum n_{i,t}$ in equation (6) into a linear combination of $n_{i,t}$. This will define some intrinsic order in the hubs, like e.g. the antiferromagnetic order in the 2-site hubs [8]. Generally, one can choose an arbitrary polynomial of the occupation numbers of the sites belonging to the hub as the argument of the Heaviside function in (6), thus defining arbitrary patterns inside the hubs.

We suggest that the multiple transitions we report here are ubiquitous and can be found in other percolation models [9-11]. Thus, one can easily show the existence of cascades of transitions in the ordinary site percolation problem. Let us take, for example, a square lattice with the probability *p* of each site to be filled, and coarse-grain it by introducing square hubs with *s* nodes and the hub fillings $v_{s,f} = \vartheta\left(\sum_{i \in hub} n_i - f\right)$. The probability of $v_{s,f} = 1$ can be readily calculated as

$$P(v_{s,f} = 1) = \sum_{k=f}^{s} \binom{s}{k} p^k (1-p)^{s-k}$$

Thus we obtain the coarse-grained square lattice built on the hubs with the new probability of a site to be filled $P(v_{s,f} = 1)$. For the critical points of the hub percolation $p_{s,f}$ we get the equation

$$\sum_{k=f}^{s} \binom{s}{k} p_{s,f}^k (1-p_{s,f})^{s-k} = p_c$$

where $p_c$ = 0.593 is the percolation threshold for a square lattice. In particular, we obtain $p_{s,s} = p_c^{1/s}$ and $p_{s,1} = 1 - (1-p_c)^{1/s}$. Note that $p_{s,s} > p_c$ while $p_{s,1} < p_c$, the latter case we can interpret as appearance of precursor percolation.

In similar manner, one can show the appearance of inner structures in many others percolation models [9-11]. Therefore, the transitions in which various percolative backbones emerge can be seen as a generic feature of percolation.

## 6. Discussion and conclusion

The main result we report in this paper is the (presumably) infinite cascades of geometric phase transitions inside the percolative phases of three kinetic models. Note that a cascade of multiple transitions is something, which we know can happen in several systems. Multiple continuous and discontinuous percolation phase transitions were reported recently in several complex network models [2, 12-16]. We should note that multiple transitions found in the networks differ from the transitions found in the present study, as the former are related to the singularities of a single order parameter, while emerging of the backbones is described by distinct order parameters.

The paradigmatic quantum Hall effect provides a well-known example of a cascade of quantum phase transitions [17, 18]. The devil's staircase of the thermal commensurate-incommensurate transitions occurs in the 3D ANNNI model [19] or in its more complicated version with the competitive in-plane interactions [20]. The experimental observations of such staircase are reported for the latter case [21].

The majority of the known multiple transitions have local variables (or their Fourier transforms) as the order parameters (OP), which are easily identified. The percolation transitions we studied belong to the other class having essentially nonlocal OPs. The singularities related to these geometric transitions have no effect on all (or almost all) local variables. This explains why we found the cascades of new transitions in the models which have been studied for at least few decades. The previous studies were mainly concentrated on the local OP for absorbing-active transition, i.e. the average site filling and its correlation functions [6, 7]. However, the study of such nonlocal variable as the survival probability in the DP model [22] revealed the signatures of multiple transitions. Numerical data of the distribution of its complex zeros show that they form a multitude of curves the ends of which tend to the real axis $p$ with the growth of a sample. This indicates the appearance of real singularities in the thermodynamic limit. The results presented in [22] are insufficient to decide to which points these curves tend exactly. However, it is quite possible that some of these limiting points are the critical points of the geometrical transitions considered here.

The situation when the system with a local transition has also the nonlocal ones is not new. The notorious example is the geometric site-percolation transition of the same-sign spins in the configuration patterns of the Ising model [23, 24] which have no effect on the local thermodynamic variables. A similar situation occurs for the Coniglio-Klein percolation order in the Ising model [25] with clusters formed by the Fortuin-Kastelein random bonds [26] on the aligned spin configurations. These clusters are designed to have the bond-percolation transition with the Ising singularities at $h$=0 [25, 24] but in addition they manifest a percolation transition along the so-called Kertesz line in $h$-$T$ plane [27, 24]. The latter has no manifestation in the thermodynamic properties of the Ising model, which has an analytical free energy at $h \neq 0$ [28]. However, the Kertesz line manifests itself in the global properties of the Ising model. According to the results [29, 30] the series expansions of the partition function of the Ising model demonstrate different convergence radii on two sides of the Kertesz line.

The system of the independent Ising spins in a magnetic field gives a simple illustration of the percolation transition completely decoupled from the local variables. When placed on some lattice with a specific set of bonds between sites, this system exhibits the site-percolation transition for the clusters of, say, up spins, while its thermodynamic (i.e. local) properties are analytic in all $h$-$T$ plane. Indeed, by introducing the occupation numbers $n_r = (1 + S_r)/2$ the featureless Gibbs distribution function of the spins can be written as the site percolation distribution function:

$$\frac{e^{\beta H \sum_r S_r}}{(2\cosh \beta H)^N} = p_+^{\sum_r n_r} (1-p_+)^{N - \sum_r n_r} , \quad p_+ \equiv \frac{e^{\beta H}}{2\cosh \beta H}$$

The distribution function on the r.h.s. of the above equation is the probability of a configuration where each site is filled independently with the probability $p_+$. It predicts divergent connected cluster of up spins (percolation) at $p_+ > p_c$, where the critical value $p_c$ is determined by the type of the lattice considered [9].

Some recent studies show that simple spin models with analytic thermodynamic parameters can exhibit other types of hidden nonlocal transitions. For example, the antiferromagnetic Ising chain in a field undergoes the transition between phases with different asymptotes of the string correlation functions [31].

These examples show that local properties of the model encoded in the partition function may not feel the presence of nonlocal transition. However, in the case of nontrivial distribution function whether the nonlocal percolation transition decouples completely from the local variables is a subtle question. In more complex systems with interaction the percolative geometric transitions can manifest themselves via crossovers of some local observables. For example, at small *h* the Kertesz line coincides with the line where the maximum of Ising magnetic susceptibility occurs [25]. Another example of such crossover was found in [8] in the growth of the average length of the "antiferromagnetic" clusters near the critical line where the hidden nonlocal order parameter, similar to the one given by equation (2), appears. Note that in the context of quantum condensed matter physics there were recent studies relating crossovers to specific critical points where nonlocal OPs (quantized topological phases and/or topological numbers) change. This was reported for the BEC-BCS crossover [32] and for several models of quantum chains and ladders [33-35]. Hence, it is natural to suggest that some crossovers of local variables are the indicators of nonlocal transitions.

Quantum phase transitions with nonlocal (hidden) topological orders can be characterized in terms of string OPs introduced first by den Nijs and Rommelse [36]. Such order parameters are found in many low-dimensional and/or frustrated systems, topological insulators/superconductors exhibiting various exotic quantum liquid states and transitions between them [17, 18]. In view of certain similarity between the quantum string OPs and those defined for the DP (2) we expect that cascades of topological transitions can be also found in many quantum systems using approaches similar to those we have employed.

In some cases it is possible to translate the hidden order into the local Landau framework, like, e.g. for the Kitaev model via transformations from spins to Majorana fermions and then to new dual spins which can manifest conventional long-ranged order [37]. However, there is no general recipe how to do it. The present results shows that the local Landau paradigm is implicitly preserved, since the scaling form of singularities (of DP universality class) implies strongly the existence of mappings of nonlocal theory exemplified by equations (1)-(3) onto local Landau-Ginzburg actions near corresponding transition points. However, now it is not clear how these mappings can be actually realized. This is an important direction for future work.

### Acknowledgments

We thank M. Herman for careful reading of the manuscript and helpful comments. We acknowledge support from the Laurentian University Research Fund (LURF) (G.Y.C.) and from the Southern Federal University grant # 213.01-2014/011-ВГ (P.N.T.). The Shared Hierarchical Academic Research Computing Network (SHARCNET) and Compute/Calcul Canada generously provided facilities for carrying out numerical calculations.